\documentclass[12pt,aps,eqsecnum,nofootinbib,floatfix]{revtex4}
\usepackage{amsmath}
\usepackage{amsfonts}
\usepackage{amssymb}
\usepackage{color}
\usepackage{graphicx}
\usepackage{mathrsfs}

\def\be{\begin{equation}}
\def\ee{\end{equation}}
\def\ba{\begin{eqnarray}}
\def\ea{\end{eqnarray}}
\def\no{\nonumber}

\definecolor{dyellow}{rgb}{1.,0.8,.0}
\definecolor{myblue}{rgb}{.1,.1,.7}
\definecolor{dcyan}{rgb}{.0,.6,.6}
\definecolor{dmagenta}{rgb}{0.6,0.0,0.6}
\definecolor{brown}{rgb}{0.6,0.2,0.}
\definecolor{darkblue}{rgb}{.0,.0,0.5}
\definecolor{darkred}{rgb}{0.75,0.0,0.0}
\definecolor{orange}{rgb}{1.,.6,.0}
\definecolor{dorange}{rgb}{0.8,.4,.0}
\definecolor{darkgreen}{rgb}{0.0,0.6,0.0}
\definecolor{purple}{rgb}{.4,.0,.4}
\definecolor{lightgrey}{rgb}{0.7, 0.7, 0.7}
\definecolor{grey}{rgb}{0.4, 0.4, 0.4}

\newcommand\btd{\raise 2pt
\hbox{$\hat\bigtriangledown$}\hskip 1.5pt}
\newcommand\bt{\raise 2pt
\hbox{$\bigtriangledown$}\hskip 1.5pt}

\newcommand{\omits}[1]{}

\begin{document}

\title{Existence condition and phase transition of Reissner-Nordstr\"{o}m-de
Sitter black hole}

\author{Meng-Sen Ma$^{a,b}$\footnote{Email: mengsenma@gmail.com}, Hui-Hua Zhao$^{a,b}$\footnote{Email:kietemap@126.com},
Li-Chun Zhang$^{b}$, Ren Zhao$^{a,b}$\footnote{Email: zhao2969@sina.com}}

\medskip

\affiliation{\footnotesize $^a$ Department of Physics, Shanxi Datong
University,  Datong 037009, China\\
$^b$Institute of Theoretical Physics, Shanxi Datong University,
Datong 037009, China}

\begin{abstract}

After introducing the connection between the black hole horizon and
the cosmological horizon, we discuss the thermodynamic properties of
Reissner-Nordstrom-de Sitter (RN-dS) spacetime. We present the
condition under which RN-dS black hole can exist. Employing
Ehrenfest' classification we conclude that the phase transition of
RN-dS black hole is the second-order one. The position of the phase
transition point is irrelevant to the electric charge of the system.
It only depends on the ratio of the black hole horizon and the
cosmological horizon.

\textbf{keywords}:Reissner-Nordstrom-de Sitter black hole, critical
phenomena, the second order phase transition
\end{abstract}


\maketitle

\tableofcontents
\bigskip


\section{Introduction}

 Black holes supply an ideal arena to research and test kinds of quantum
gravity\cite{Lu1}. On the one hand, black holes are the solutions of
classical gravity (GR) , thus are classical systems. On the other
hand, black holes can also be a kind of quantum system, in which
thermodynamics plays an important
role\cite{JDB1,JDB2,JDB3,BCH,Hawking1,Hawking2}. The entropy,
temperature and the holographic properties of black holes are
essentially related to quantum mechanics. Although the statistical
explanation of the thermodynamic states of black holes is lacked
yet, the relevant studies on the properties of black hole
thermodynamics still received a lot of attentions, such as
Hawking-Page phase transition\cite{Hawking3}, the critical
phenomena. More interestingly, it is found for some black holes
there exist similar phase transition and critical behaviors to the
van der Waals-Maxwell system\cite{chamblin,chamblin1}.

Recently, the idea of including the variation of the cosmological
constant $\Lambda $ in the first law of black hole thermodynamics
has attained increasing attention
\cite{Dolan1,Dolan2,Dolan3,Dolan4,Cvetic,RBM,RBM2,LYX1,Cai1,Hendi,Belhaj,Zhao1,Bagher,Spallucci,Kim,MMS,Liu,Mann,Wang,Zhaoliu,Cadoni}.
Matching the thermodynamic quantities with the ones in usual
thermodynamic system, the critical behavior of black holes can be
investigated and the phase diagram like the van der Waals
vapor-liquid system can be obtained. This helps to further
understand black hole entropy, temperature, heat capacity, et.al,
and it is also very important to improve the self-consistent
geometric theory of thermodynamics of black hole

The black holes mentioned above must be stable in the appropriate
range of parameters. Only in this case the corresponding
relationship between the black hole and the usual thermodynamic
system can be true. The black holes , like Schwarzschild black hole,
are thermodynamic unstable. To investigate their thermodynamic
properties, some methods are proposed, for example black branes
thermodynamics\cite{Vaidya,APL,Lu2,Lu3,Lu4}. The black holes in de
Sitter space usually possess not only the black hole horizon, but
the cosmological horizon. The horizons all have thermal radiation,
thus different temperatures. Therefore black holes in de Sitter
spacetime are thermodynamic unstable. The thermodynamic quantities
on the black hole horizon and the cosmological horizon all satisfy
the first law of thermodynamics, moreover the corresponding
entropies both fulfill the area formula\cite{RBM,Cai2,Sekiwa}. In
recent years the studies on the thermodynamic properties of de
Sitter space have aroused wide
concern\cite{RBM,Cai2,Sekiwa,Urano,Zhang,Myung}. In the era of
inflation the universe lies in a quasi-de Sitter space. The
cosmological constant corresponds to vacuum energy and is usually
considered as a candidate of dark energy. The accelerating universe
will evolve into another de Sitter phase. In order to construct the
entire history of evolution of the universe, we should have a clear
perspective to the classical and quantum properties of de Sitter
space\cite{RBM,Sekiwa,Cai3,SB}. Firstly, we anticipate the entropy
should satisfy the Nernst theorem\cite{Urano,Zhang,Cai4}. Secondly,
after introducing the connection between the black hole horizon and
the cosmological horizon, we want to know whether the thermodynamic
quantities in de Sitter space fulfill the conditions of
thermodynamic stability, whether in de Sitter space there exists
similar phase transition and critical phenomena like in AdS space?
Hence constructing a self-consistent relation between the
thermodynamic quantities in de Sitter space is worth studying.

The thermodynamic quantities corresponding to the black hole horizon
and the cosmological horizon are all functions of the mass $M$,
electric charge $Q$ and the cosmological constant $\Lambda $.
Therefore the two pairs of thermodynamic quantities are not
independent each other. Taking their relations into account is very
important to study the thermodynamic properties in de Sitter space.
In \cite{Zhao2} we set the position $r_c$ of the cosmological
horizon to be invariant and studied the phase transition of RN-dS
black hole. Based on the results in \cite{Zhao2}, in this paper we
investigate the critical behaviors of the effective thermodynamic
quantities of RN-dS spacetime when  the  cosmological horizon is
variable. According to Ehrenfest's classification for phase
transition of thermodynamic system, we find that what happens in
RN-dS black hole is the second-order phase transition.

The paper is arranged as follows: In Sec.2 we first review the RN-dS
spacetime and give the thermodynamic quantities corresponding to the
two horizons. After considering the connection between the two
horizons we introduce the effective temperature, effective pressure
and effective potential. In Sec. 3 phase transition in charged dS
black hole spacetime is investigated. We discuss the relation
between the effective pressure and the effective volume in RN-dS
spacetime and analyze its critical phenomena. We will analyze the
nature of the phase transition using Ehrenfest's equations in Sec.
4. Finally, the paper ends with a brief conclusion. (we use the
units $G=\hbar =k_B =c=1)$

\section{The effective thermodynamic quantities of RN-dS spacetime}

The line element of the RN-dS black holes is given by\cite{Cai2}
\begin{equation}
\label{eq1} ds^2=-f(r)dt^2+f^{-1}dr^2+r^2d\Omega ^2,
\end{equation}
where
\begin{equation}
\label{eq2} f(r)=1-\frac{2M}{r}+\frac{Q^2}{r^2}-\frac{\Lambda
}{3}r^2,
\end{equation}
The above geometry possesses three horizons: the black hole Cauchy
horizon locates at $r=r_- $, the black hole event horizon (BEH)
locates at $r=r_+ $ and the cosmological event horizon (CEH) locates
at $r=r_c $, where $r_c
>r_+ >r_- $; the only real, positive zeroes of $f(r)=0$. After considering
the connection between the black hole horizon and the cosmological
horizon, the thermodynamic relation in RN-dS spacetime is provide in
\cite{Zhao2}
\begin{equation}
\label{eq3} dM=T_{eff} dS-P_{eff} dV+\varphi _{eff} dQ,
\end{equation}
Where the effective temperature is \ba\label{eq2.4} T_{eff}
&=&\frac{(1+x-2x^2+x^3+x^4)}{4\pi r_c
x(1+x)(1+x+x^2)}-\frac{Q^2}{4\pi r_c^3 x^3(x+1)(x^2+x+1)}
\left( {1+x+x^2-2x^3+x^4+x^5+x^6} \right)\no\\
&=&\frac{B_1 }{4\pi r_c x(x+1)} -\frac{Q^2B_2 }{4\pi r_c^3 x^3(1+x)}
\ea The effective pressure is \ba\label{eq4}
P_{eff}&=&\frac{(1-x)(1+3x+3x^2+3x^3+x^4)}{8\pi r_c^2
x(1+x)(1+x+x^2)^2}
-\frac{Q^2(1+2x+3x^2-3x^5-2x^6-x^7)}{8\pi r_c^4 x^3(1+x)(1+x+x^2)^2} \no\\
&=& \frac{B_3 }{8\pi r_c^2 x(1+x)} -\frac{Q^2B_4 }{8\pi r_c^4
x^3(1+x)} \ea And the effective electric potential
\begin{equation}
\label{eq5} \varphi _{eff} =Q\frac{r_c^4 +2r_c^3 r_+ +2r_c^2 r_+^2
+2r_c r_+^3 +r_+^4 }{r_c r_+ (r_c +r_+ )(r_c^2 +r_+^2 +r_c r_+ )}
=Q\frac{1+x+x^2+x^3}{r_c x(1+x+x^2)}.
\end{equation}
The thermodynamic volume in RN-dS spacetime is \cite{Vaidya,SB}
\begin{equation}
\label{eq6} V=\frac{4\pi }{3}\left( {r_c^3 -r_+^3 } \right).
\end{equation}
The entropy of RN-dS system is\cite{Zhao3}
\begin{equation}
\label{eq7} S=S_+ +S_c .
\end{equation}
where \ba &&B_1 =\frac{1+x-2x^2+x^3+x^4}{1+x+x^2}, \quad
B_2 =\frac{1+x+x^2-2x^3+x^4+x^5+x^6}{1+x+x^2},\no\\
&&B_3 =\frac{1+2x-2x^4-x^5}{(1+x+x^2)^2}, \quad ~~~~~ B_4
=\frac{1+2x+3x^2-3x^5-2x^6-x^7}{(1+x+x^2)^2}. \ea and $x:=r_+ /r_c
$, $0<x<1$. $S_+ $ and $S_c $ are the entropies which correspond to
the black hole horizon and the cosmological horizon respectively.
When $Q=0$, the thermodynamic equation (\ref{eq3}) will return back
to the known result in \cite{Urano}. When the relation $r_c^2 x^2B_1
>Q^2B_2 $ is fulfilled, $T_{eff} >0$ is always satisfied.

The effective quantities defined in (\ref{eq2.4}),(\ref{eq4}) and
(\ref{eq5}) and the thermodynamic equation (\ref{eq3}) are
self-consistent. When $x\to 1$, namely the two horizons tend to
coincide,
\begin{equation}
\label{eq9} P_{eff} \to 0,\quad T_{eff} \to \frac{1}{12\pi r_c^3
}\left( {r_c^2 -2Q^2} \right),
\end{equation}
In this state, because
\[
Q^2=r_+ r_c \left( {1-\frac{r_c^2 +r_c r_+ +r_+^2 }{3}\Lambda }
\right), \quad 2M=(r_c +r_+ )\left( {1-\frac{r_c^2 +r_+^2
}{3}\Lambda } \right)
\]
\begin{equation}
\label{eq10} 2M=\frac{(r_c +r_+ )}{r_+^2 +r_c r_+ +r_c^2 }\left(
{r_c r_+ +Q^2\frac{r_c^2 +r_+^2 }{r_c r_+ }} \right).
\end{equation}
thus one can obtain
\begin{equation}
\label{eq11} Q^2=r_c^2 (1-r_c^2 \Lambda ), \quad M=r_c \left(
{1-\textstyle{2 \over 3}r_c^2 \Lambda } \right),
\end{equation}
Due to $Q^2\ge 0, ~M\ge 0$, so $r_c^2 \Lambda \le 1$. When $M^2\ge
Q^2$ is satisfied, $r_c^2 \Lambda >\frac{3}{4}$ can be derived
according to (\ref{eq11}), therefore $T_{eff} \to \frac{1}{12\pi
_c^3 }\left( {2r_c^2 \Lambda -1} \right)>0$, which fulfills the
requirement of the thermodynamic equilibrium stability. If do not
considering the connection between the two horizons and taking them
into account as independent thermodynamic systems, due to the
different radiation temperature for the two horizons, the spacetime
is not stable.

Another problem to perceive the two horizons as independent
thermodynamic systems is when the two horizons coincide, $\kappa
_{+/c} =0$ which means the temperature from the both horizons is
zero. In this case the areas of the both horizons are obviously not
zero, namely the entropies correspond to the black hole horizon and
the cosmological horizon are not zero. Therefore Nernst theorem
cannot be satisfied\cite{Zhao5,Mi}.

When the two horizons coincide, from (\ref{eq6}), the thermodynamic
volume $V\to 0$ in the RN-dS spacetime. Hence the thermodynamic
system transits from volume distribution to area distribution. The
pressure of the `` thermodynamic brane'' is zero and volume tends to
zero, but the temperature $T_{eff} $ of the `` thermodynamic brane''
is not zero, the entropy $S\to 2\pi r_c^2 $. This may explain the
problem that extremal de Sitter black holes do not satisfy Nernst
theorem.

\section{P-V criticality in RN-dS black hole spacetime}

The investigation of phase transition in thermodynamic system has
been an active subject. Recently by treating black holes as the
thermodynamic systems many works on the phase transition of black
holes has been
done\cite{Cvetic,RBM,RBM2,LYX1,Cai1,Hendi,Zhao1,Bagher,Vaidya,APL,Lu2,Lu3,Lu4,Sahay1,Sahay2,Sahay3,Kastor,RB1,RB2,RB3,RB4,RB5,RB6,RB7,RB8,Majhi,Tain}.
Moreover by comparing with the Van der Waals system the critical
behaviors of black hole system are also be
studied\cite{Cvetic,RBM,RBM2,LYX1,Cai1,Hendi,Belhaj,Zhao1,Bagher,Vaidya,APL,Lu2,Lu3,Lu4}.
However, due to the existence of two horizons in de Sitter
spacetime, two different thermodynamic systems for the two horizons
should be built. There is few research on the phase transition of
this kind of non-equilibrium system. Based on Sec. 2, we will
investigate the phase transition of RN-dS black holes. First we
compare the effective thermodynamic quantities in RN-dS black hole
with the Van der Waals equation and discuss the relation between
pressure and volume at constant temperature. Then we will analyze
the nature of the phase transition using Ehrenfest's equations.

Comparing with the Van der Waals equation
\begin{equation}
\label{eq12} \left( {P+\frac{a}{v^2}} \right)(v-\tilde {b})=kT,
\end{equation}
Here, $v=V/N$ is the specific volume of the fluid, $P$ its pressure,
$T$ its temperature, and $k$ is the Boltzmann constant. From
(\ref{eq12}) one can plot the $P-v$ curves at constant $T$. The
critical temperature, critical pressure and the critical specific
volume can be determined according to the first and second
derivatives.

To compare with Van der Waals equation we set $P_{eff} \to P, ~v\to
v$ and discuss the phase transition and critical phenomena at
constant $Q$.

Substituting (\ref{eq2.4}) into (\ref{eq4}), one can derive
\begin{equation}
\label{eq13} P_{eff} =T_{eff} \frac{B_4 }{2r_c B_2 }+\frac{B_2 B_3
-B_1 B_4 }{8\pi r_c^2 x(1+x)B_2 },
\end{equation}
$x$ is a dimensionless parameter. Employing (\ref{eq13}) and the
thermodynamic volume (\ref{eq6}) in the RN-dS spacetime, by
dimensional analysis\cite{Cvetic} we can set the specific volume to
be
\begin{equation}
\label{eq14} v=r_c (1-x).
\end{equation}
The critical point occur when the two equations set up at the same
time:
\begin{equation}
\label{eq15} \left( {\frac{\partial P_{eff} }{\partial v}}
\right)_{T_{eff} } =0, \quad \left( {\frac{\partial ^2P_{eff}
}{\partial v^2}} \right)_{T_{eff} } =0,
\end{equation}
\begin{equation}
\label{eq16} \left( {\frac{\partial P_{eff} }{\partial v}}
\right)_{T_{eff} } =\left( {\frac{\frac{\partial (T_{eff} ,P_{eff}
)}{\partial (x,r_c )}}{\frac{\partial (T_{eff} ,v)}{\partial (x,r_c
)}}} \right) =\left( {\frac{\left( {\frac{\partial P_{eff}
}{\partial r_c }} \right)_x \left( {\frac{\partial T_{eff}
}{\partial x}} \right)_{r_c } -\left( {\frac{\partial P_{eff}
}{\partial x}} \right)_{r_c } \left( {\frac{\partial T_{eff}
}{\partial r_c }} \right)_x }{r_c \left( {\frac{\partial T_{eff}
}{\partial r_c }} \right)_x +(1-x)\left( {\frac{\partial T_{eff}
}{\partial x}} \right)_{r_c } }} \right).
\end{equation}
letting
\begin{equation}
\label{eq17} \left( {\frac{\left( {\frac{\partial P_{eff} }{\partial
r_c }} \right)_x \left( {\frac{\partial T_{eff} }{\partial x}}
\right)_{r_c } -\left( {\frac{\partial P_{eff} }{\partial x}}
\right)_{r_c } \left( {\frac{\partial T_{eff} }{\partial r_c }}
\right)_x }{r_c \left( {\frac{\partial T_{eff} }{\partial r_c }}
\right)_x +(1-x)\left( {\frac{\partial T_{eff} }{\partial x}}
\right)_{r_c } }} \right) =f(x,r_c ),
\end{equation}
From which we can derive
\begin{equation}
\label{eq18} \left( {\frac{\partial ^2P_{eff} }{\partial v^2}}
\right)_{T_{eff} } = \left( {\frac{\left( {\frac{\partial
f}{\partial r_c }} \right)_x \left( {\frac{\partial T_{eff}
}{\partial x}} \right)_{r_c } -\left( {\frac{\partial f}{\partial
x}} \right)_{r_c } \left( {\frac{\partial T_{eff} }{\partial r_c }}
\right)_x }{r_c \left( {\frac{\partial T_{eff} }{\partial r_c }}
\right)_x +(1-x)\left( {\frac{\partial T_{eff} }{\partial x}}
\right)_{r_c } }} \right) =0.
\end{equation}

One can choose different $Q$ and calculate the critical values of
the effective thermodynamic quantities numerically. We give them in
Table I. In Fig.1 we plot the $P_{eff} -v$ curves when $Q=1,3,10$
respectively.
\begin{table}[htb]
\centering
\begin{tabular}{c c c c c c}
\hline\hline $Q$ & $x^c$  & $r_c^c$ &$v^c$ &$T_{eff}^c$ &
$P_{eff}^c$  \\ [0.05ex] \hline
1 & 0.732216 & 3.5062 & 0.938907 &0.00801475 & 0.00060544 \\
3 & 0.732216 & 10.5186 & 2.81672 & 0.00267158 & 0.0000672711 \\
10 & 0.732216 & 35.062 & 9.38907 & 0.000801475 & $6.0554\times 10^{-6}$ \\[0.05ex]
\hline
\end{tabular}\label{E1}
\caption{Numerical solutions for $x^c$, $r_c^c$, $v^c$, $T_{eff}^c$
and $P_{eff}^c$ for given values of $Q=1, 3, 10$ respectively. }
\label{tab1}
\end{table}

\begin{figure}[ht]
\centering
\includegraphics[scale=0.4,keepaspectratio]{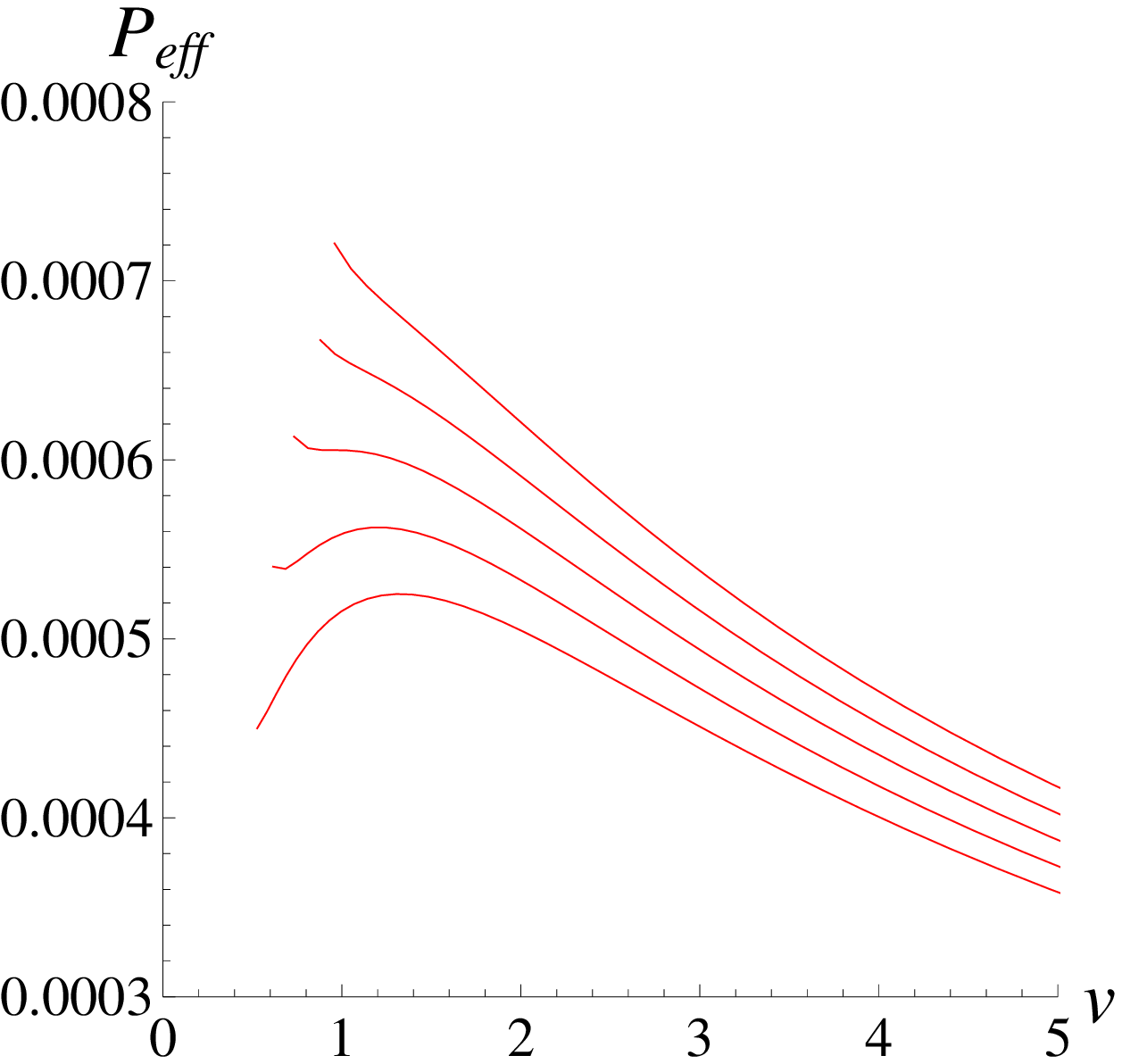}
\includegraphics[scale=0.4,keepaspectratio]{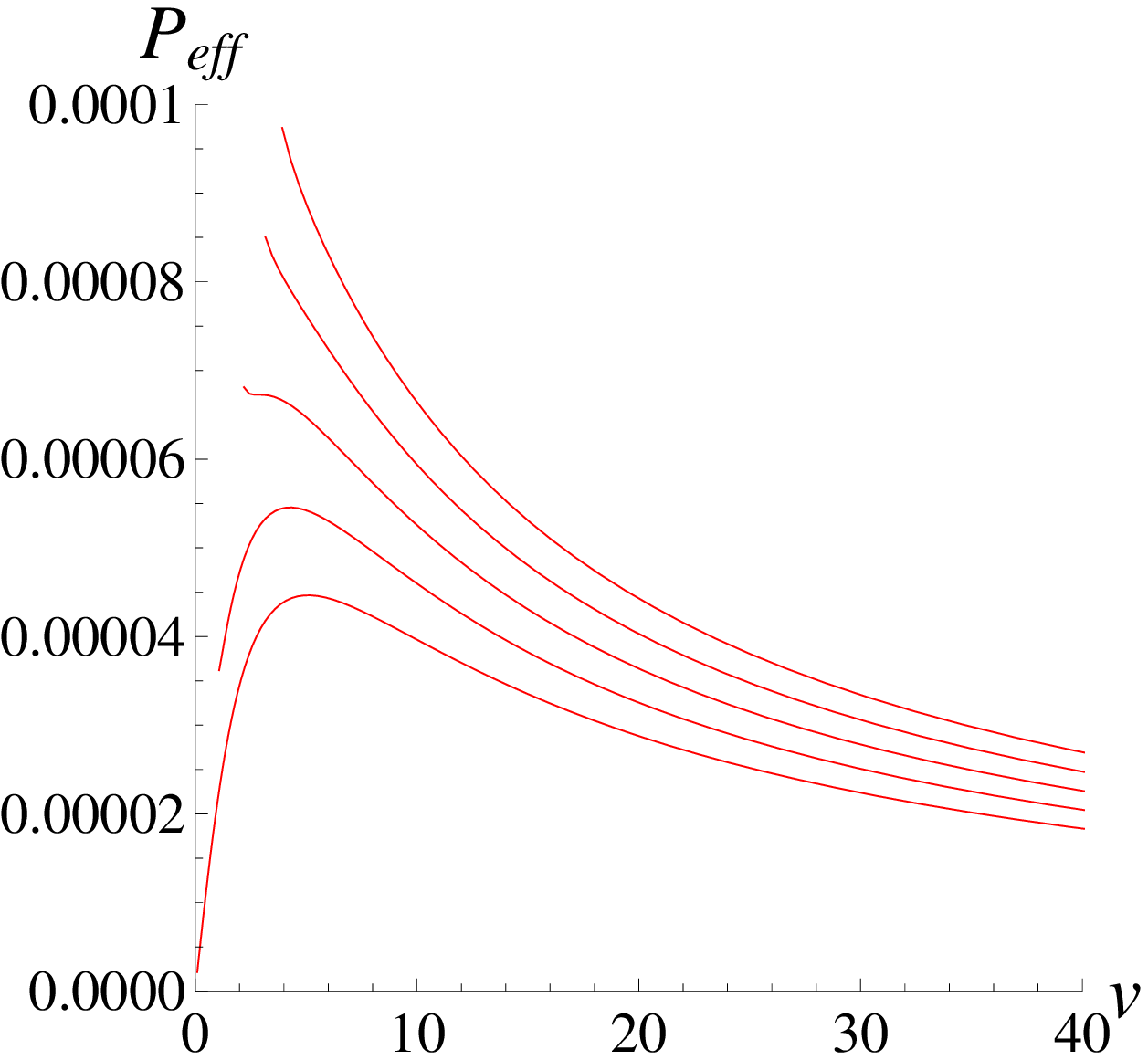}
\includegraphics[scale=0.4,keepaspectratio]{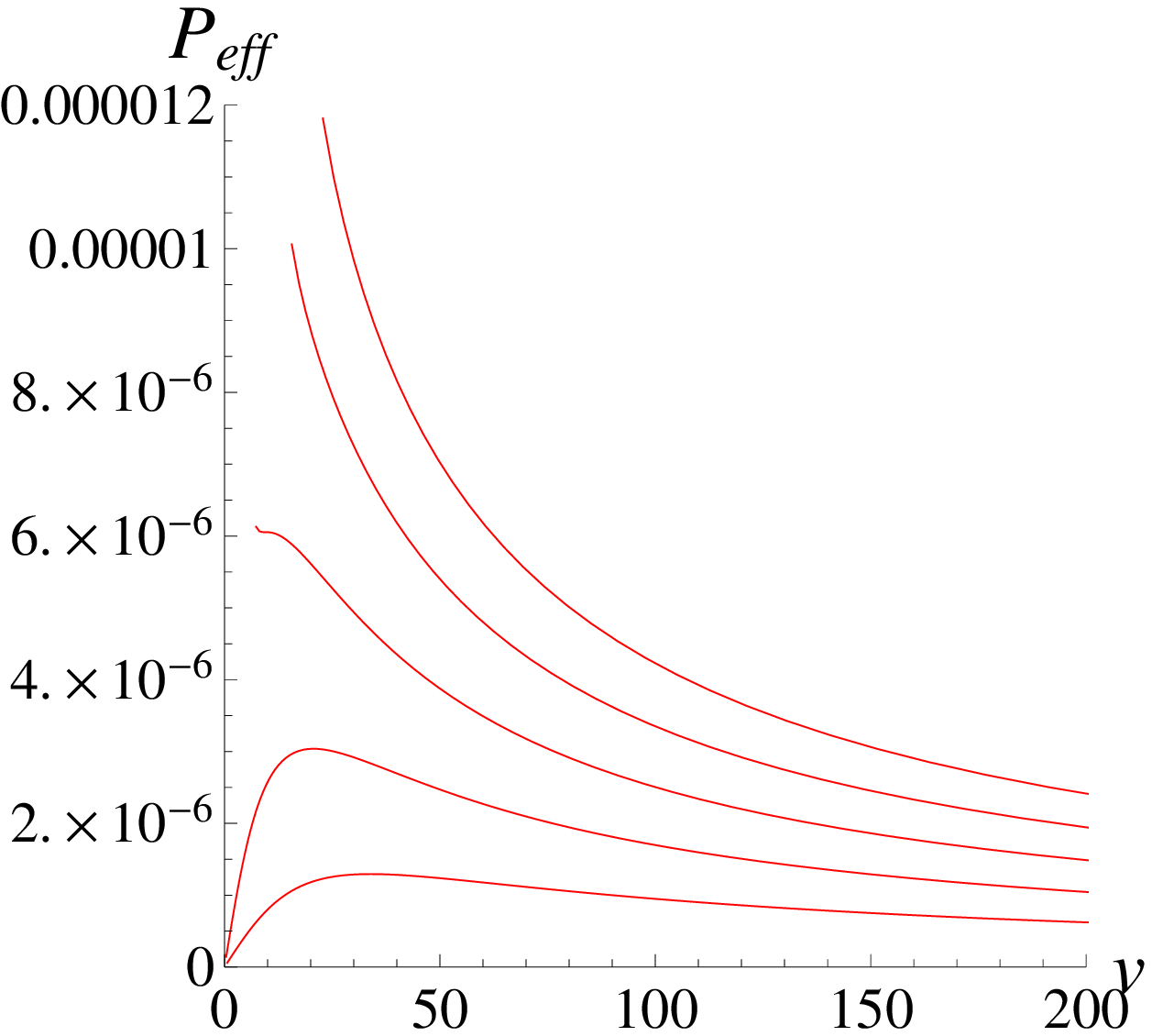}
\caption[]{\it The $p-v$ curves for $Q=1, 3, 10$ respectively. From
top to the bottom the curves correspond to the effective temperature
$T_{eff}^c +0.0004$˙$T_{eff}^c +0.0002$˙$T_{eff}^c $˙$T_{eff}^c
-0.0002$ and $T_{eff}^c -0.0004$.} \label{figurepv}
\end{figure}

From the above calculation, we find that the value of $x$ is
$x^c=0.732216$ at the critical point in the RN-dS black hole, which
is independent of the electric charge. The critical temperature is
inversely proportional to the electric charge, $T_{eff}^c
=\frac{0.00281475}{Q}$. The critical cosmological horizon $r_c^c $
is proportional to the charge, $r_c^c =10.5186Q$, and the critical
specific volume $v^c$ is also related to the electric charge,
$v^c=0.938907Q$.

From Fig.1, when $T_{eff} >T_{eff}^c $,RN-dS system satisfies the
stability condition $\left( {\frac{\partial P}{\partial v}}
\right)_{T_{eff} } <0$; when $T_{eff} <T_{eff}^c $, for larger value
of $v$ RN-dS system satisfies the stability condition $\left(
{\frac{\partial P}{\partial v}} \right)_{T_{eff} } <0$, however, for
smaller value of $v$ RN-dS system does not satisfy the stability
condition. Thus those states may not exist in nature.

\section{The second-order phase transition of RN-dS spacetime}

For Van der Waals system there is no latent heat and at the critical
point the liquid-gas structure do not change suddenly. Therefore
this kind of phase transition belongs to the continuous phase
transition according to Ehrenfest's classification. Below we will
discuss the behaviors of RN-dS system near the phase transition
point.

When the chemical potential and its first derivative is continuous,
whereas the second derivative of chemical potential is
discontinuous, this kind of phase transition is called the
second-order phase transition. We can calculate the specific heat of
RN-dS system at constant pressure $C_P $, the expansion coefficient
$\beta $ and the compressibility $\kappa $ \ba\label{eq19} C_P
&=&T_{eff} \left( {\frac{\partial S}{\partial T_{eff} }}
\right)_{P_{eff}
} =-T_{eff} \frac{\partial ^2\mbox{G}}{\partial T_{eff}^2 }\no \\
&=&\pi r_c T_{eff} \left( {\frac{r_c x\left( {\frac{\partial P_{eff}
}{\partial r_c }} \right)_x -(1+x^2)\left( {\frac{\partial P_{eff}
}{\partial x}} \right)_{r_c } }{\left( {\frac{\partial T_{eff}
}{\partial x}} \right)_{r_c } \left( {\frac{\partial P_{eff}
}{\partial r_c }} \right)_x -\left( {\frac{\partial T_{eff}
}{\partial r_c }} \right)_x \left( {\frac{\partial P_{eff}
}{\partial x}} \right)_{r_c } }} \right), \ea

\ba\label{eq20} \beta &=&\frac{1}{v}\left( {\frac{\partial
v}{\partial T_{eff} }} \right)_{P_{eff} } =\frac{1}{v}\frac{\partial
^2\mu }{\partial T_{eff}
\partial P_{eff} } \no \\
&=&-\frac{1}{v}\left( {\frac{r_c \left( {\frac{\partial P_{eff}
}{\partial r_c }} \right)_x +(1-x)\left( {\frac{\partial P_{eff}
}{\partial x}} \right)_{r_c } }{\left( {\frac{\partial T_{eff}
}{\partial x}} \right)_{r_c } \left( {\frac{\partial P_{eff}
}{\partial r_c }} \right)_x -\left( {\frac{\partial T_{eff}
}{\partial r_c }} \right)_x \left( {\frac{\partial P_{eff}
}{\partial x}} \right)_{r_c } }} \right) , \ea

\ba\label{eq21} \kappa &=&-\frac{1}{v}\left( {\frac{\partial
v}{\partial P_{eff} }}
\right)_{T_{eff} } =-\frac{1}{v}\frac{\partial ^2\mu }{\partial P_{eff}^2 }\no \\
&=&-\frac{1}{v}\left( {\frac{r_c \left( {\frac{\partial T_{eff}
}{\partial r_c }} \right)_x +(1-x)\left( {\frac{\partial T_{eff}
}{\partial x}} \right)_{r_c } }{\left( {\frac{\partial P_{eff}
}{\partial r_c }} \right)_x \left( {\frac{\partial T_{eff}
}{\partial x}} \right)_{r_c } -\left( {\frac{\partial P_{eff}
}{\partial x}} \right)_{r_c } \left( {\frac{\partial T_{eff}
}{\partial r_c }} \right)_x }} \right). \ea From (\ref{eq7}), the
entropy is
\begin{equation}
\label{eq22} S=\pi r_c^2 (1+x^2).
\end{equation}

Below we will give the $\kappa-x$, $\beta -x$, $C_P -x$, $S-T$,
$G-T$ curves for the fixed values of $Q=1, 3, 10$.

\begin{figure}[ht]
\centering
\includegraphics[scale=0.4,keepaspectratio]{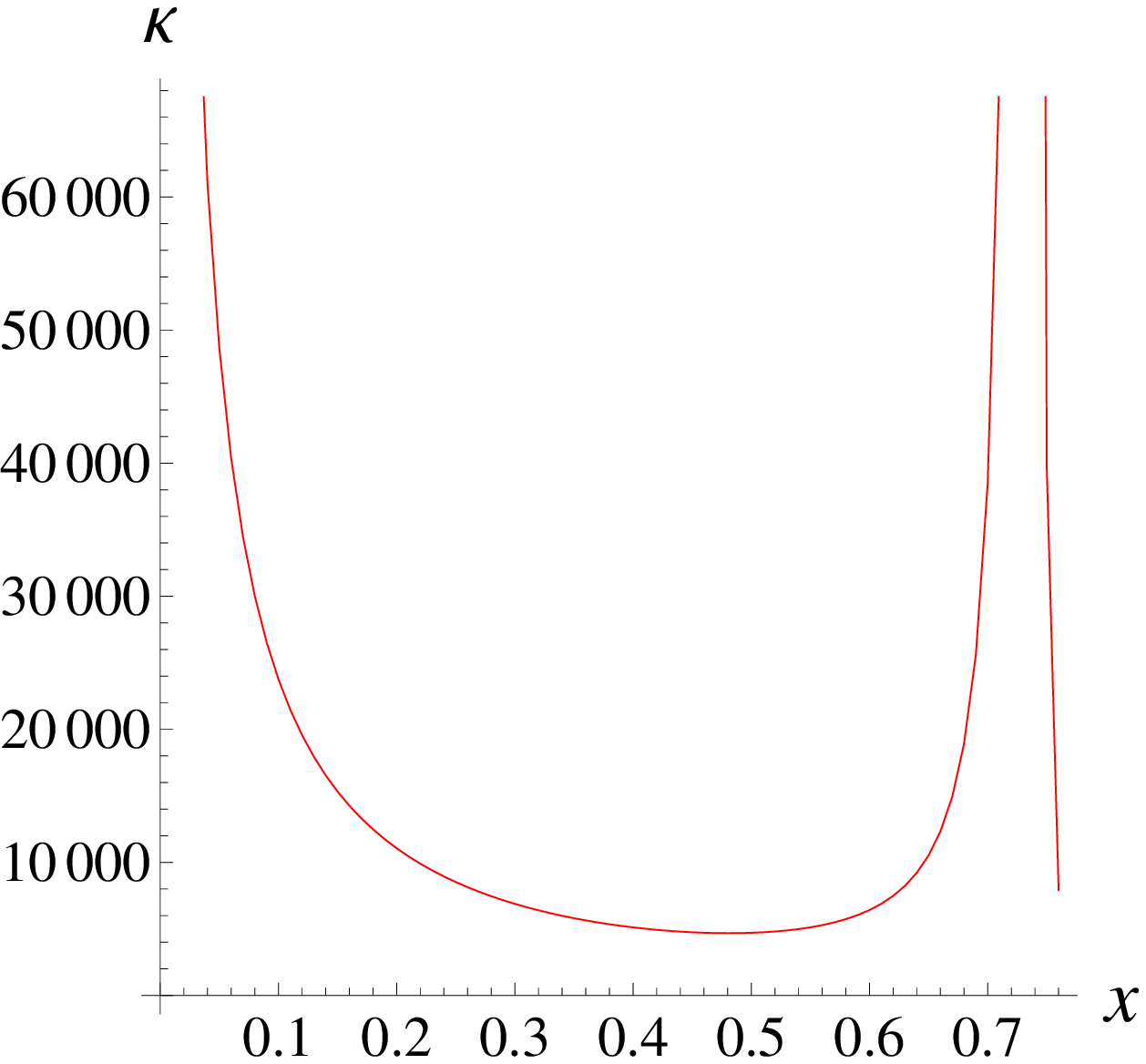}
\includegraphics[scale=0.4,keepaspectratio]{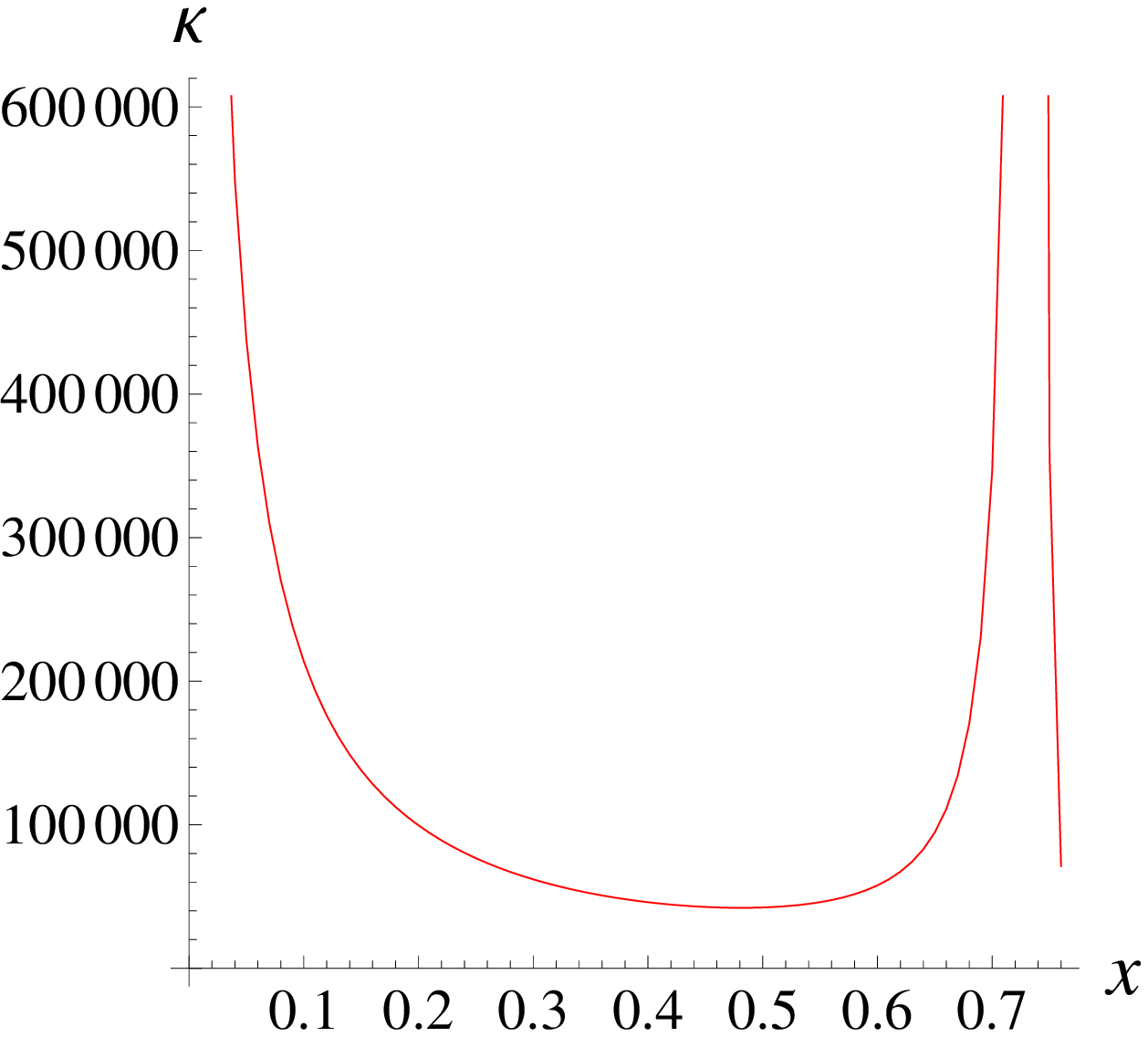}
\includegraphics[scale=0.4,keepaspectratio]{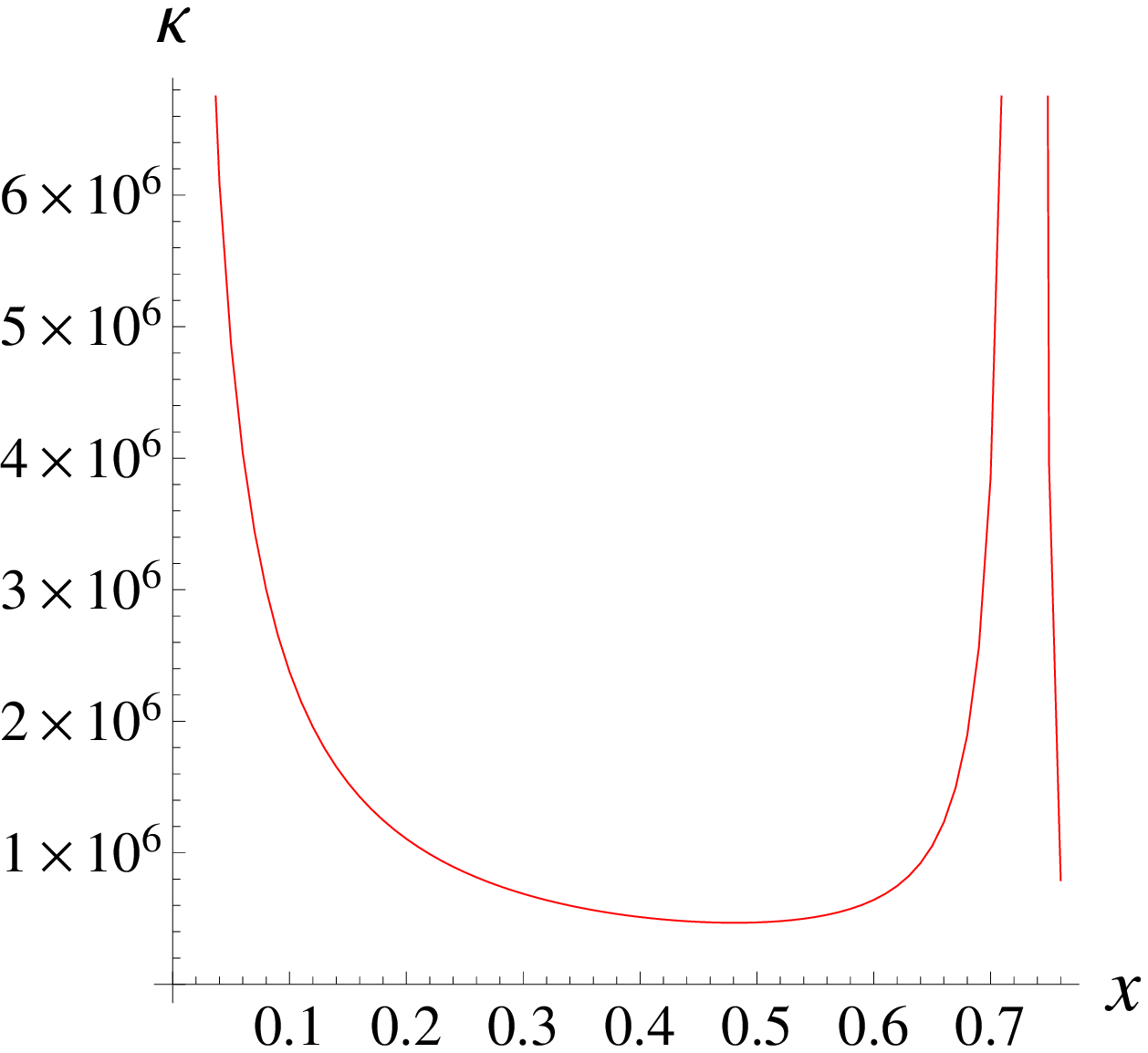}
\caption[]{\it $\kappa-x$ curves for RN-dS black hole corresponding
to the critical effective temperature  $T_{eff}^c =0.00801475$,
$T_{eff}^c =0.00267158$ and $T_{eff}^c =0.000801475$ respectively.}
 \label{figureKx}
\end{figure}

\begin{figure}[ht]
\centering
\includegraphics[scale=0.4,keepaspectratio]{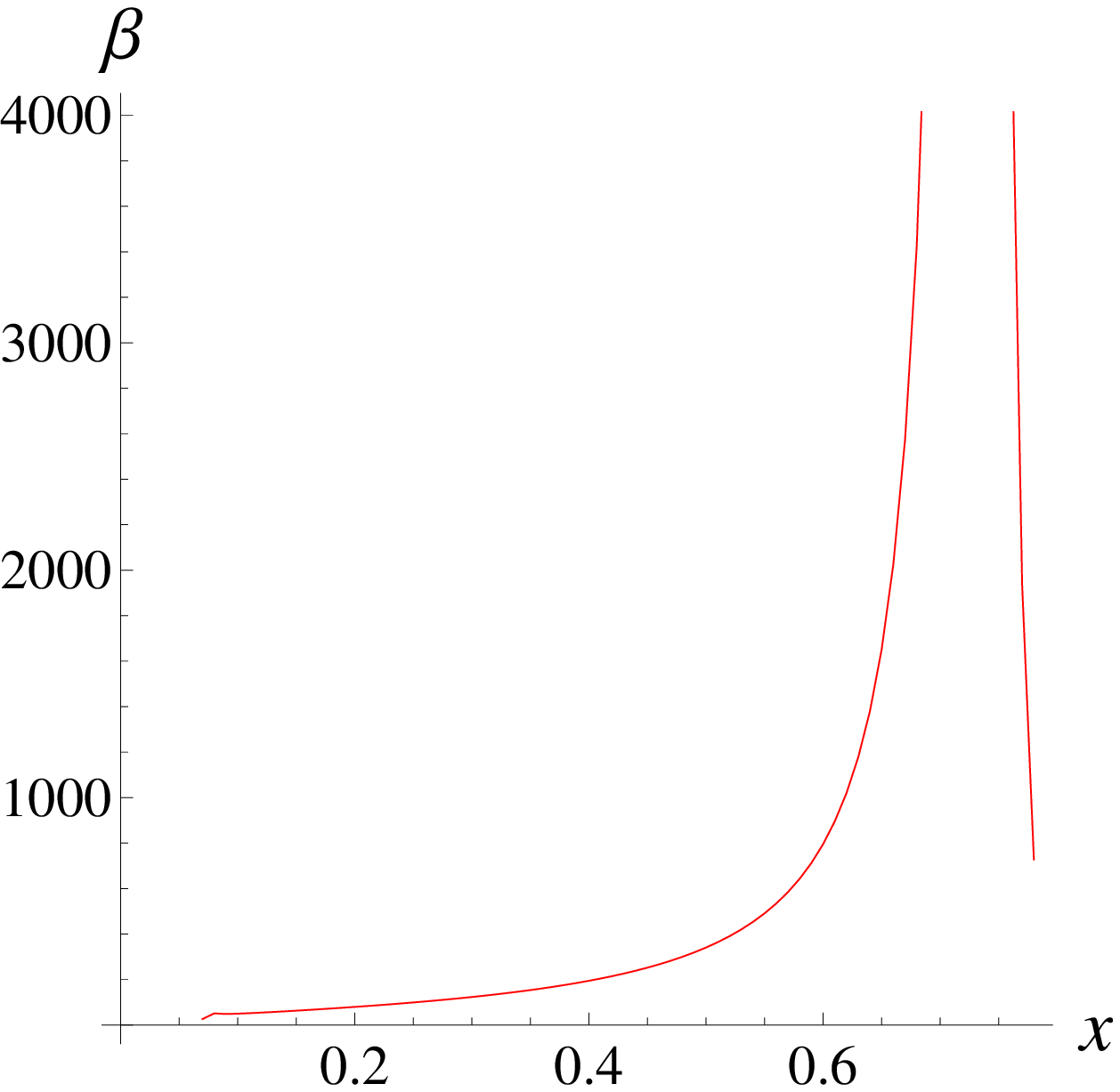}
\includegraphics[scale=0.4,keepaspectratio]{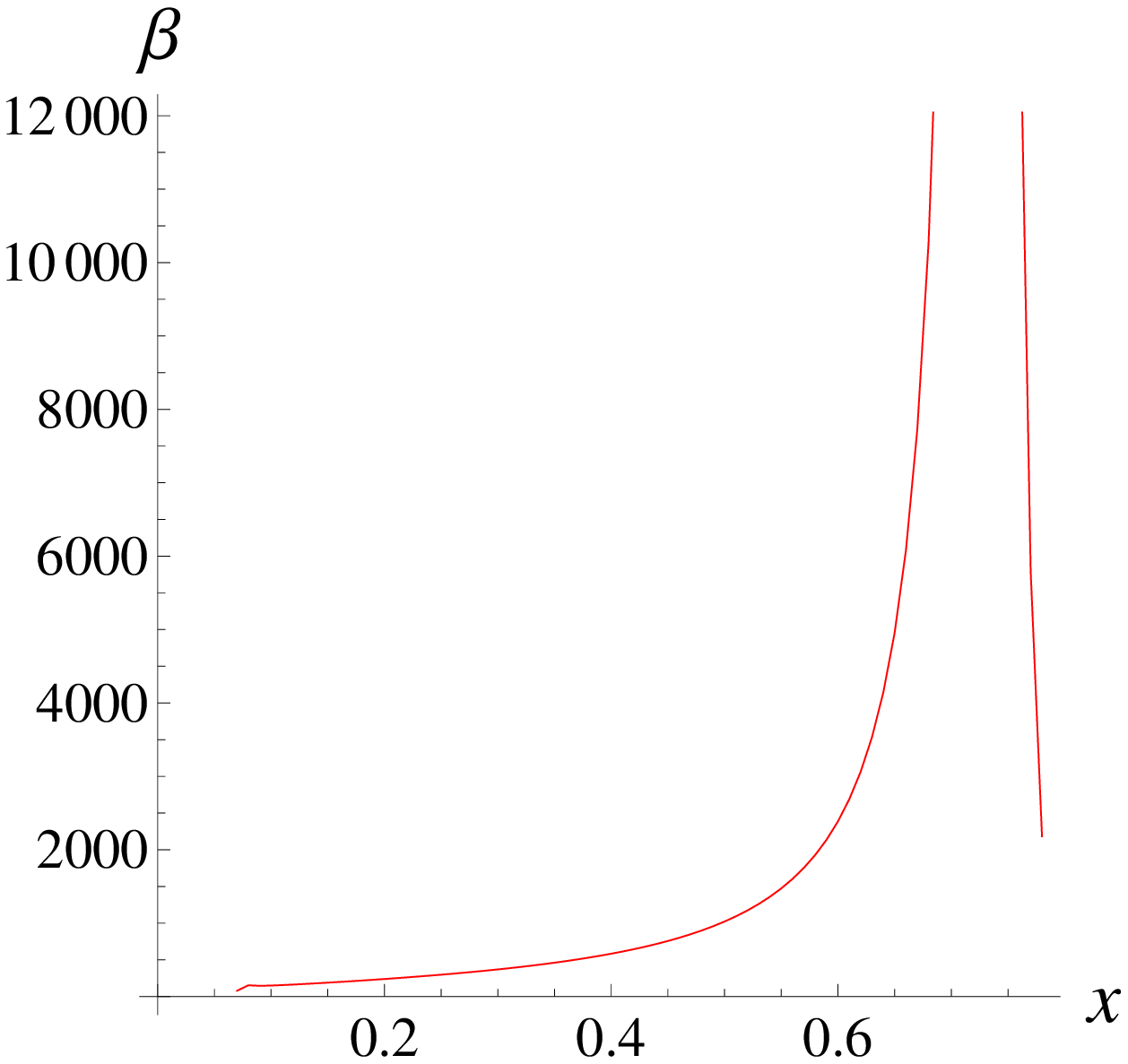}
\includegraphics[scale=0.4,keepaspectratio]{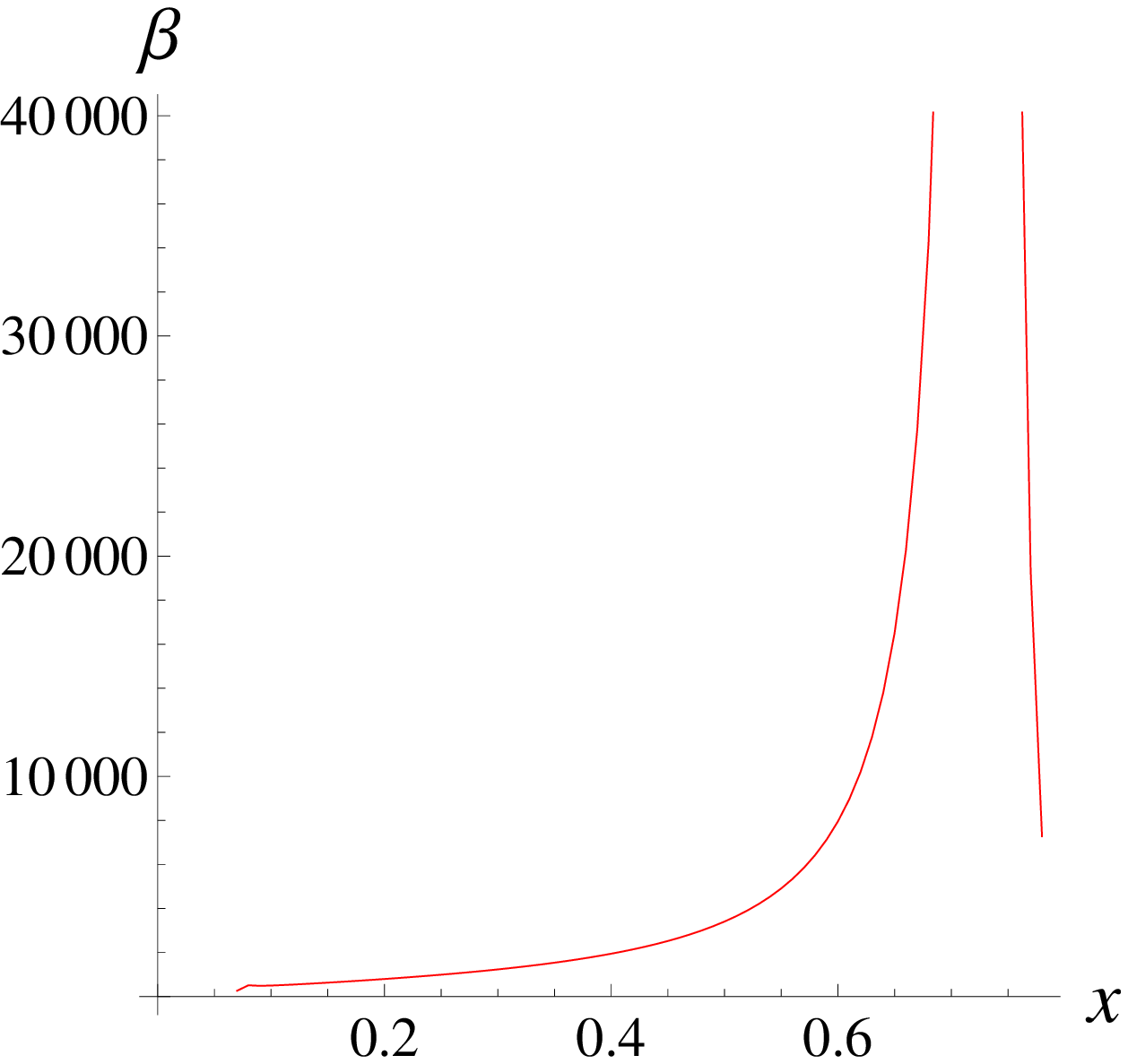}
\caption[]{\it $\beta-x$ curves for RN-dS black hole corresponding
to the critical effective pressure  $P_{eff}^c =0.00060544$,
$p_{eff}^c =0.0000672711$ and $P_{eff}^c =6.0554\times 10^{-6}$
respectively.} \label{figureBx}
\end{figure}

\begin{figure}[ht]
\centering
\includegraphics[scale=0.4,keepaspectratio]{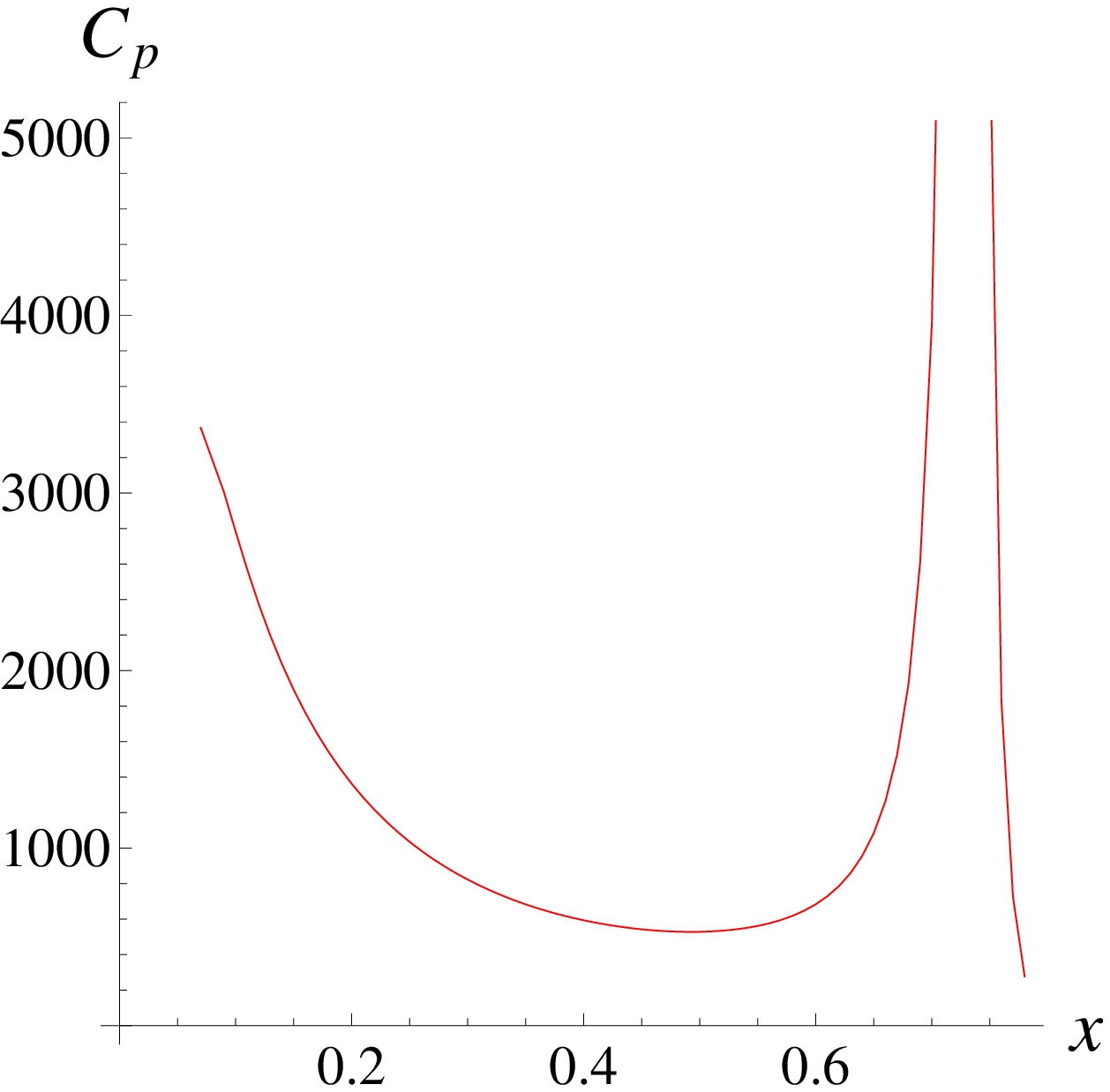}
\includegraphics[scale=0.4,keepaspectratio]{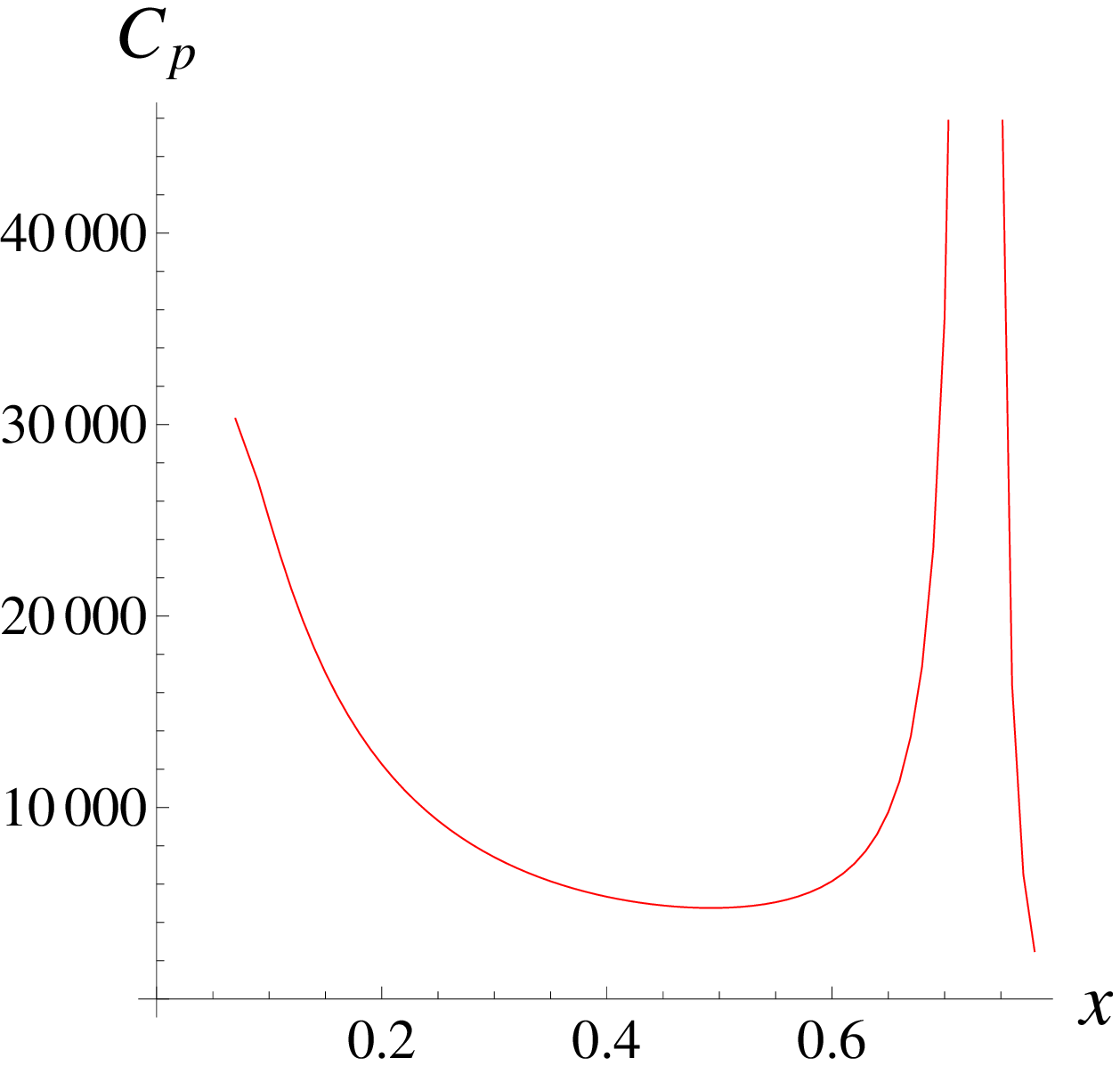}
\includegraphics[scale=0.4,keepaspectratio]{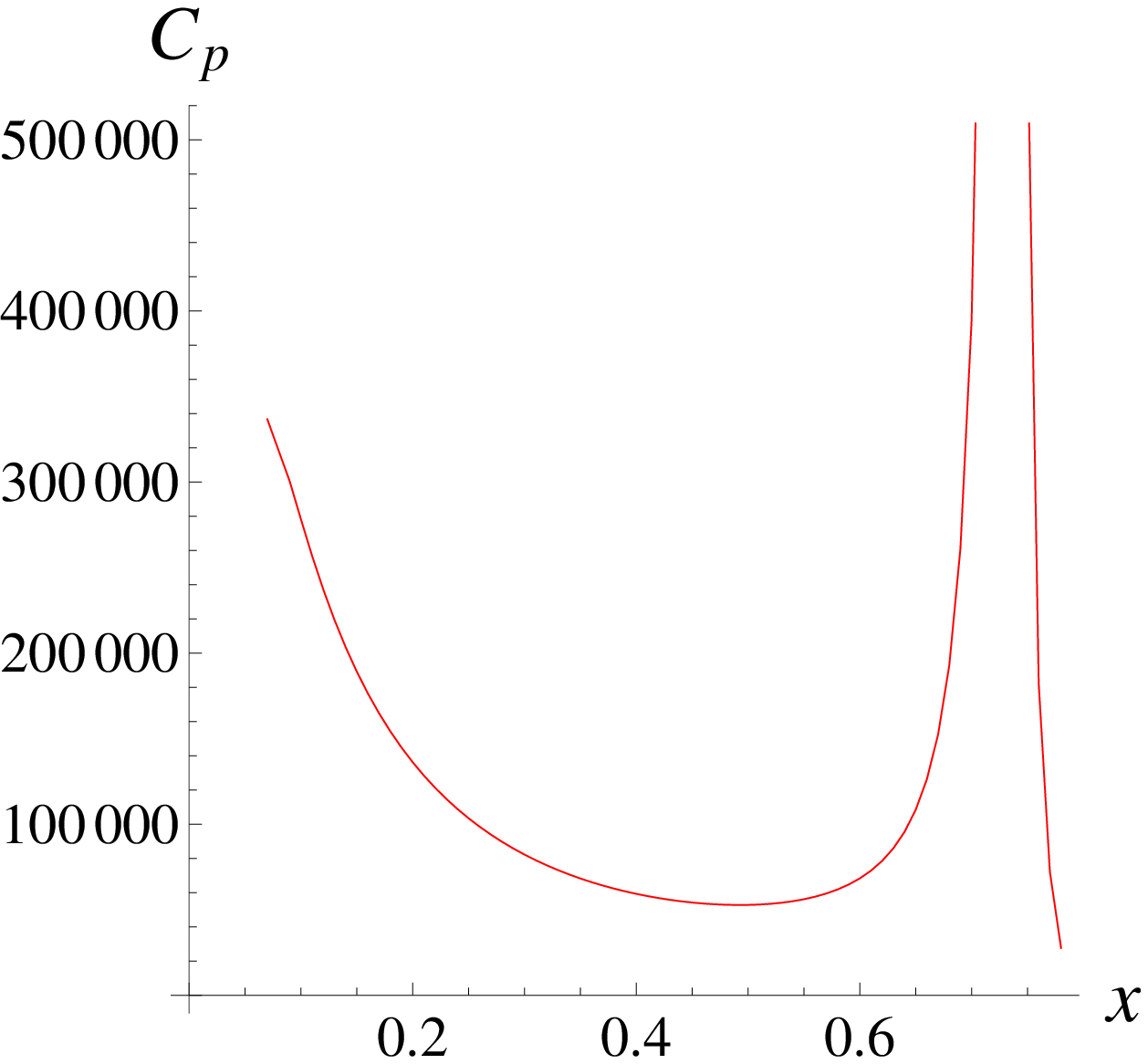}
\caption[]{\it $C_P-x$ curves for RN-dS black hole corresponding to
the critical effective pressure  $P_{eff}^c =0.00060544$, $p_{eff}^c
=0.0000672711$ and $P_{eff}^c =6.0554\times 10^{-6}$ respectively.}
\label{figureCpx}
\end{figure}

\begin{figure}[htb]
\centering
\includegraphics[scale=0.4,keepaspectratio]{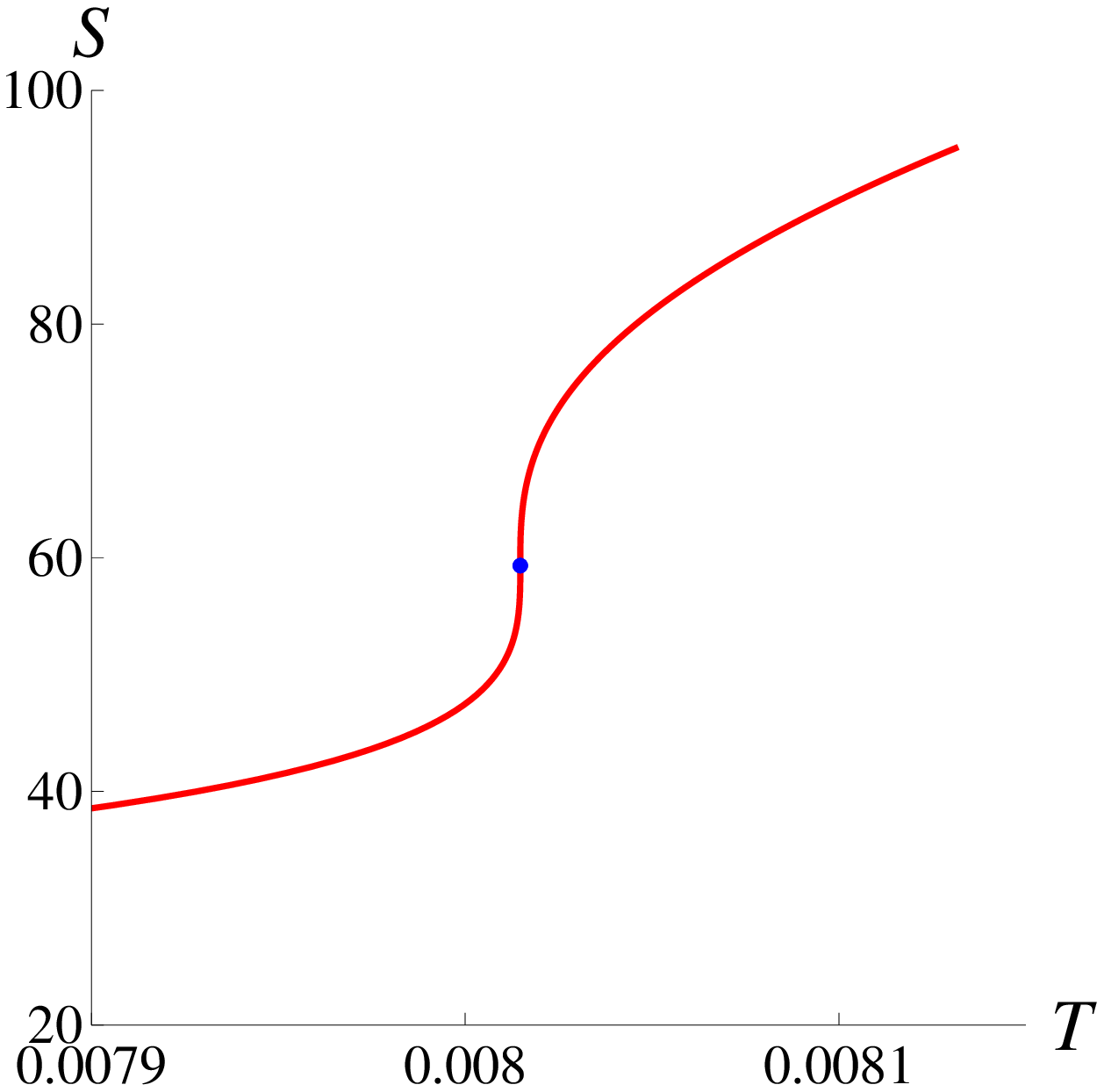}
\includegraphics[scale=0.4,keepaspectratio]{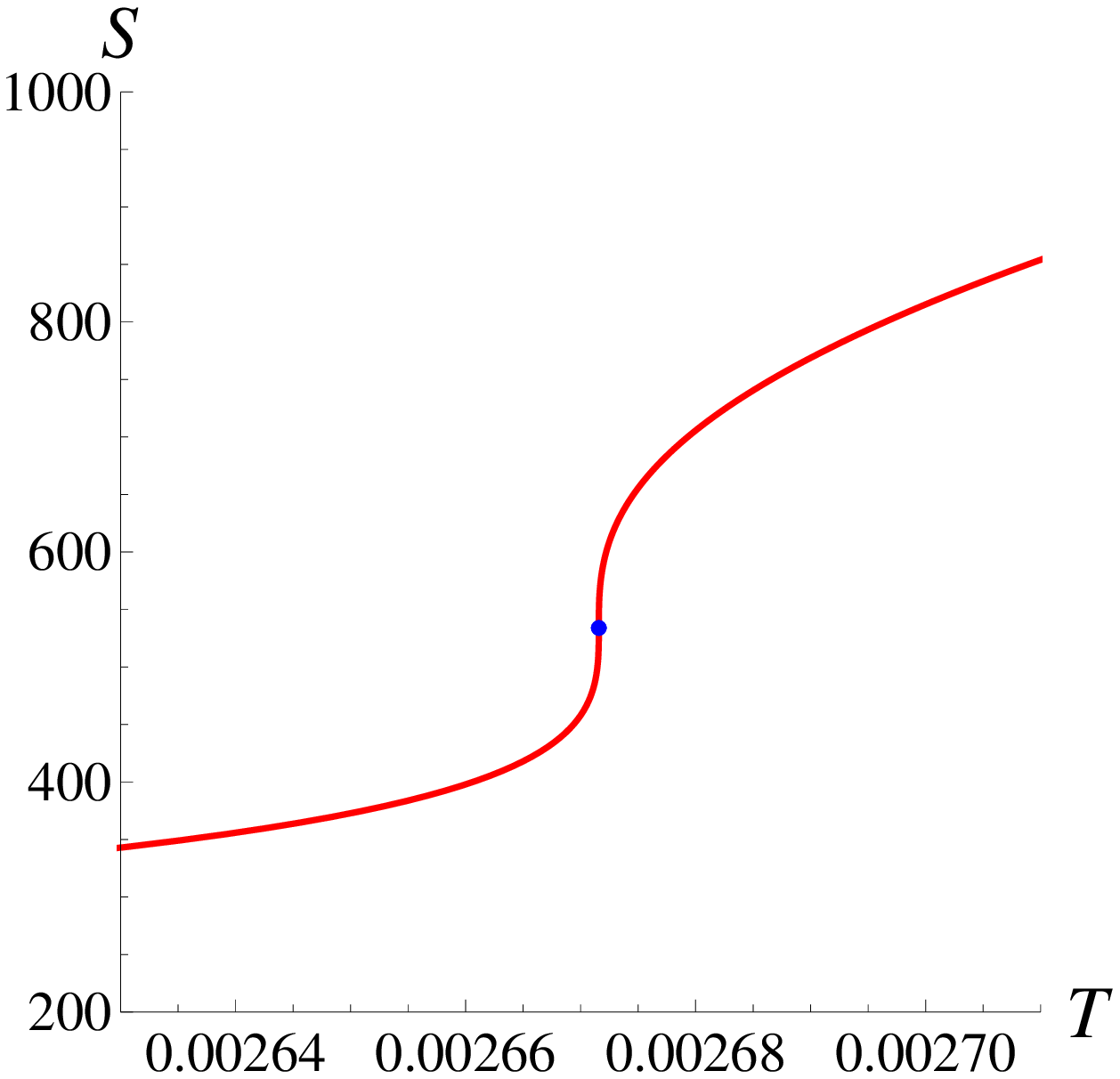}
\includegraphics[scale=0.4,keepaspectratio]{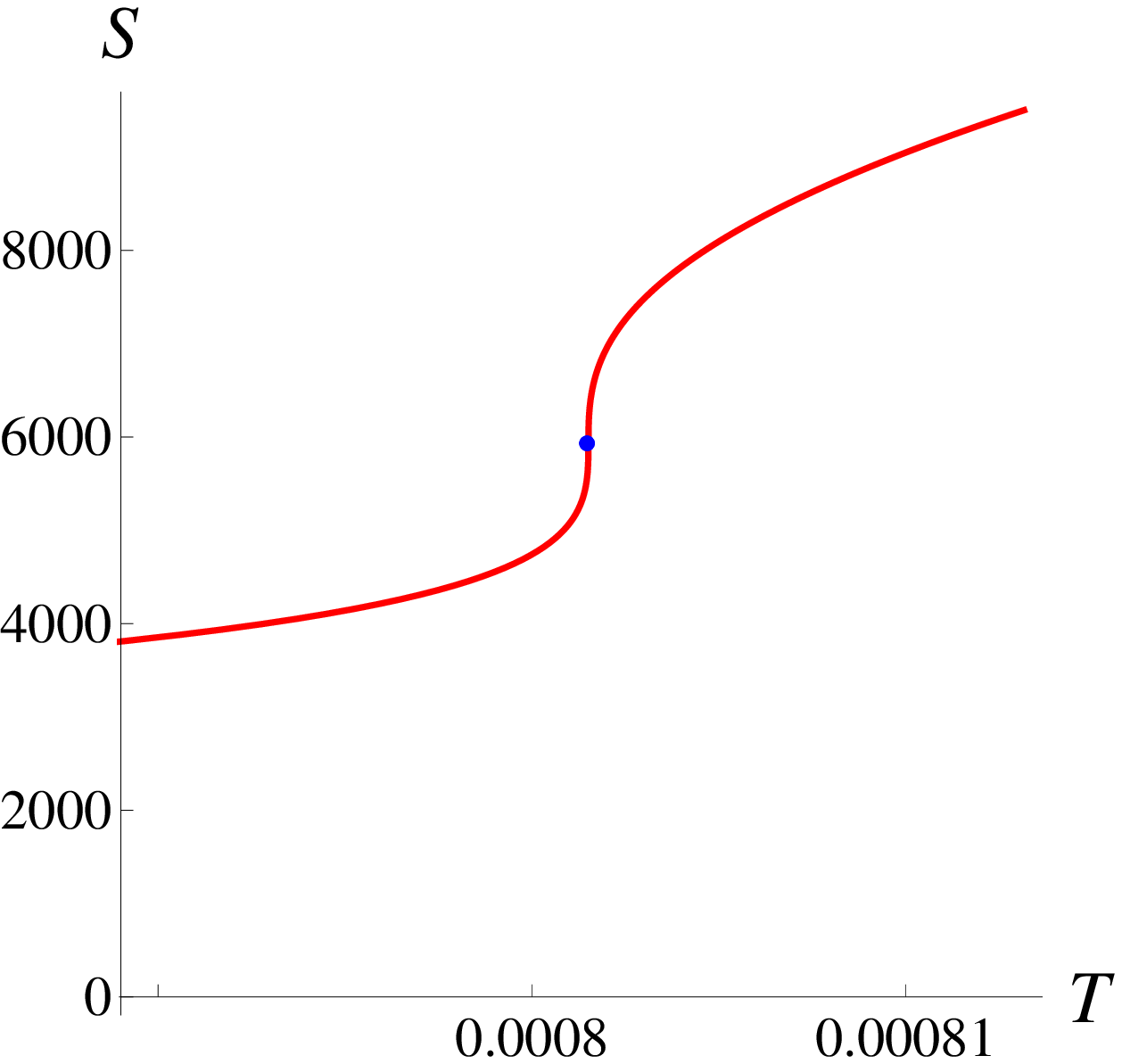}
\caption[]{\it $S-T$ curves for RN-dS black hole corresponding to
the critical effective pressure  $P_{eff}^c =0.00060544$, $p_{eff}^c
=0.0000672711$ and $P_{eff}^c =6.0554\times 10^{-6}$ respectively.}
\label{figureST}
\end{figure}

\begin{figure}[htb]
\centering
\includegraphics[scale=0.4,keepaspectratio]{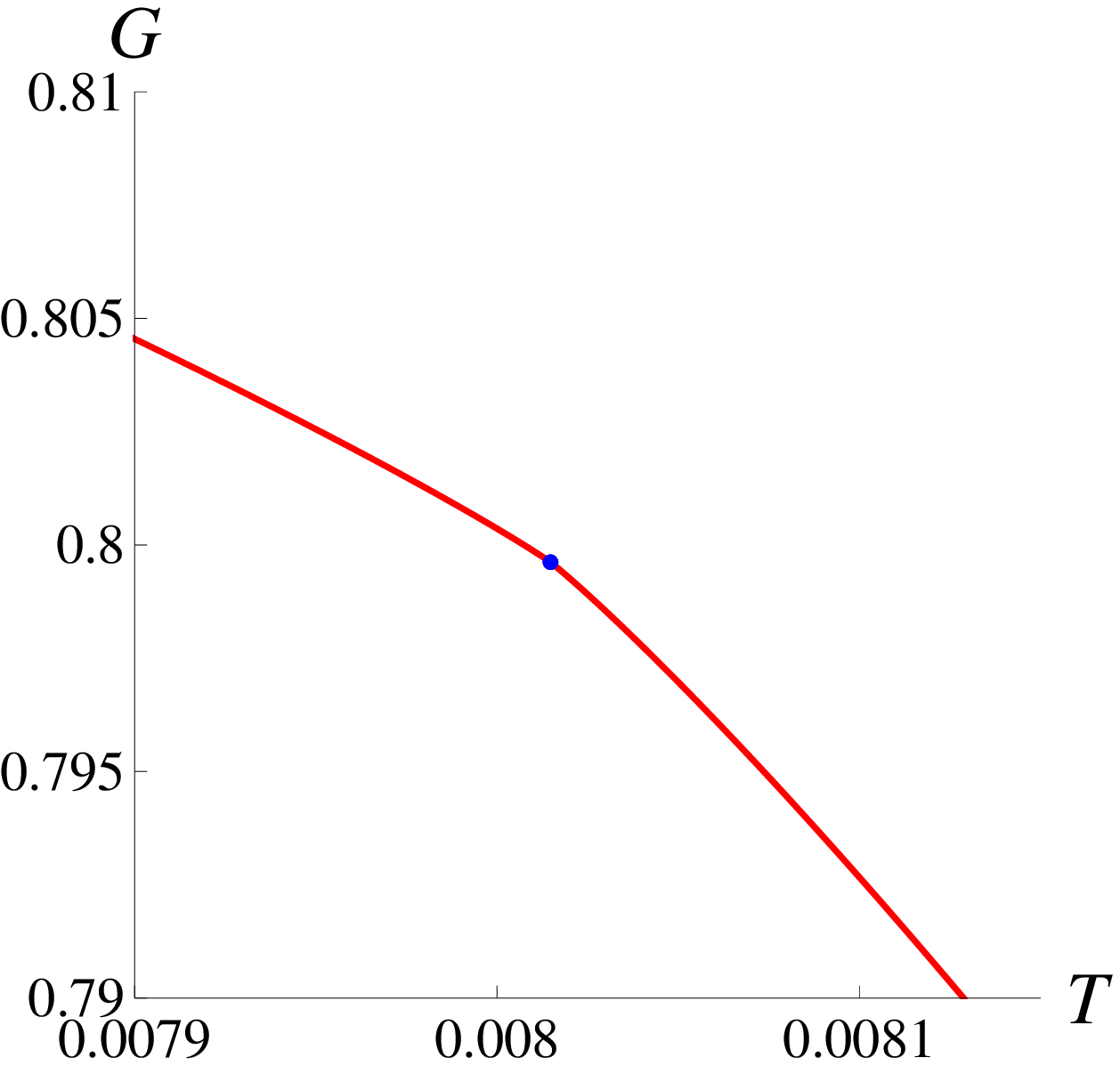}
\includegraphics[scale=0.4,keepaspectratio]{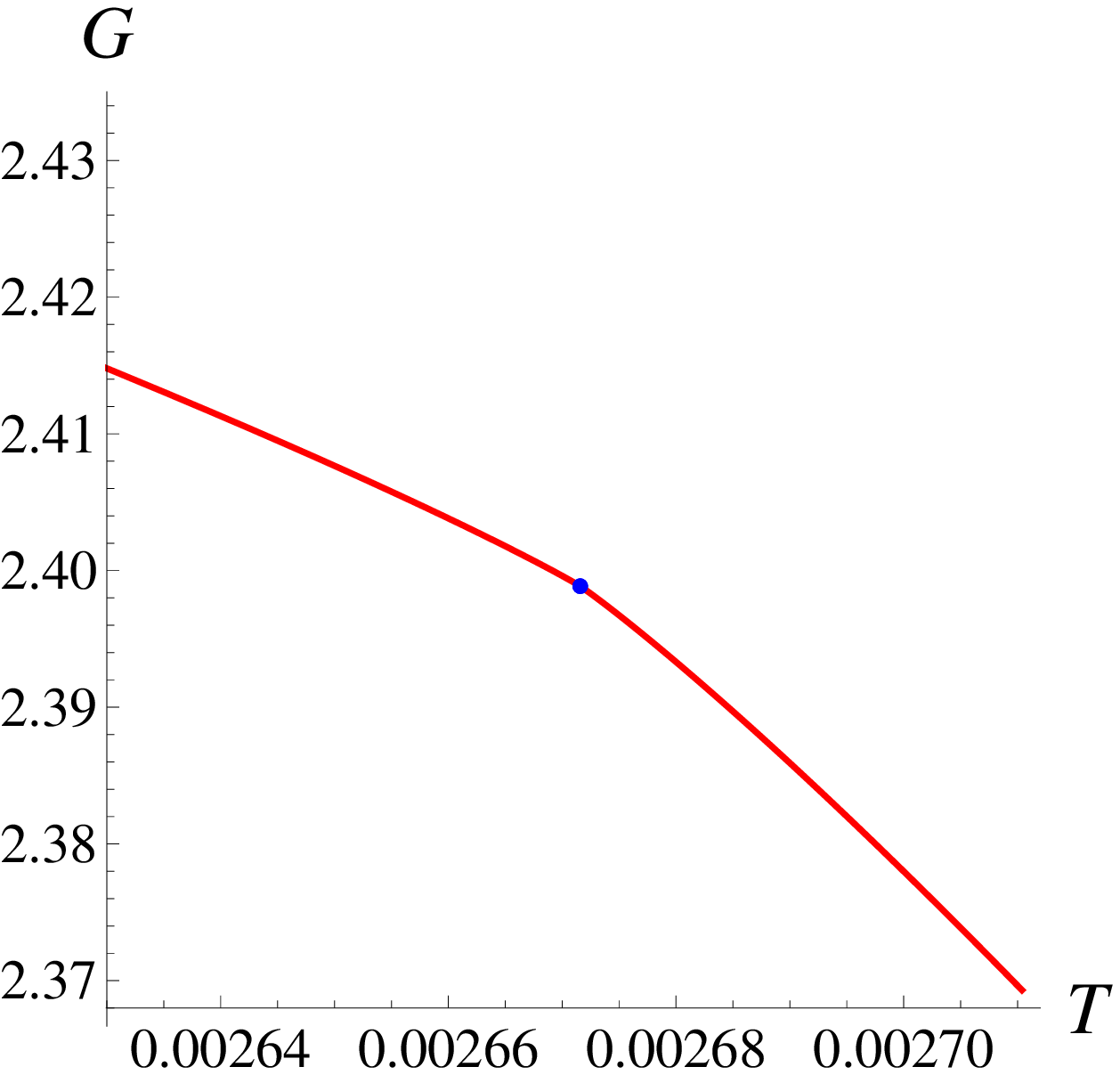}
\includegraphics[scale=0.4,keepaspectratio]{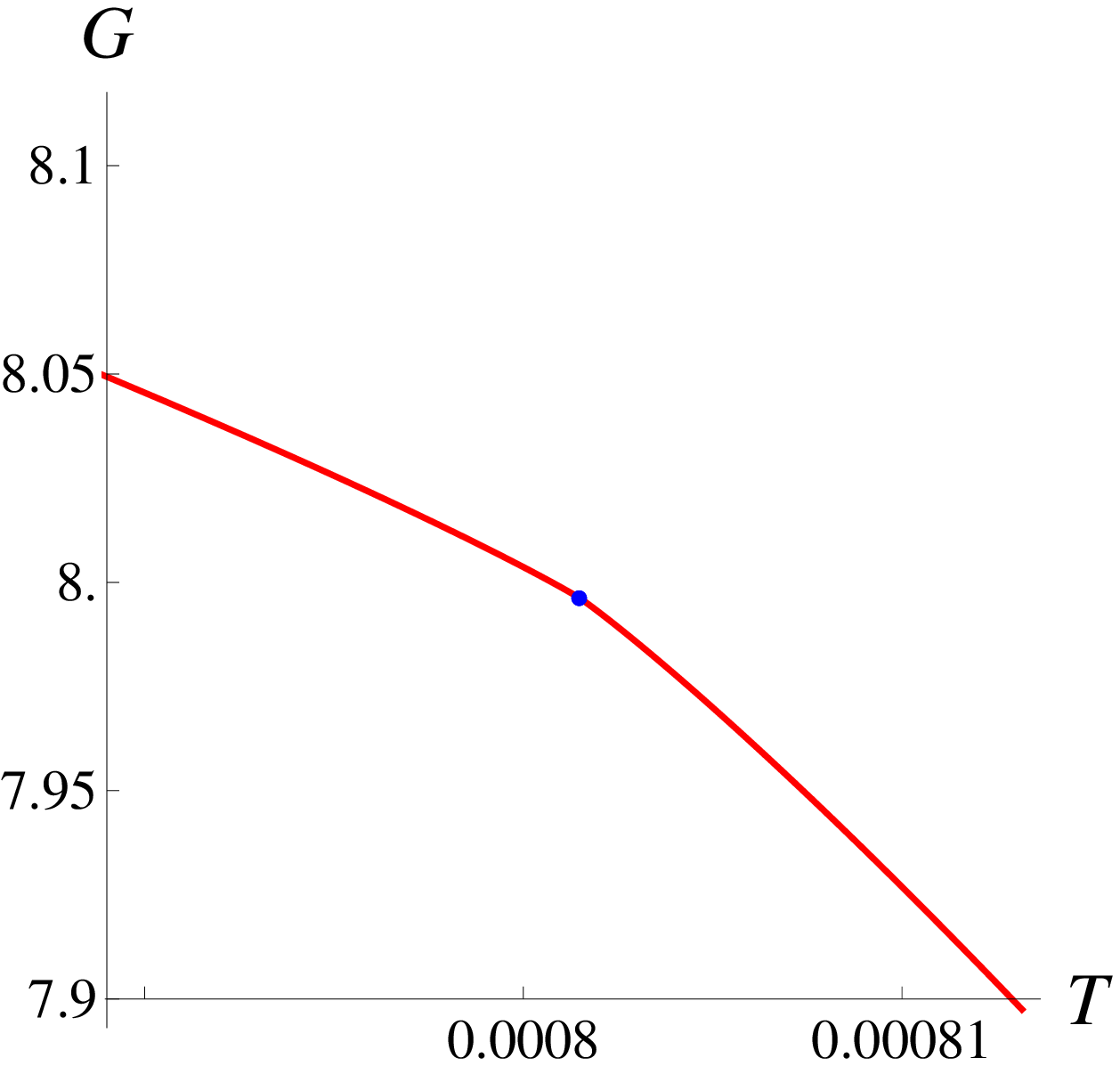}
\caption[]{\it $G-T$ curves for RN-dS black hole corresponding to
the critical effective pressure  $P_{eff}^c =0.00060544$, $p_{eff}^c
=0.0000672711$ and $P_{eff}^c =6.0554\times 10^{-6}$ respectively.
For fixed charge $Q$, the Gibbs free energy can be expressed as
$G=M-T_{eff} S-P_{eff} V$.[1109.2433; 1208.6251;1306.4516˙
1209.1707ㄛ1203.2279]} \label{figureGT}
\end{figure}

From the above figures, it can be found that the specific heat at
constant pressure, the expansion coefficient $\beta $ and the
compressibility $\kappa $ exist infinite peak. While the Gibbs
function $G$ and the entropy $S$ are both continuous at the critical
point. According to Ehrenfest, the phase transition of the RN-dS
black hole should be the second-order one.

\section{ Conclusions and Discussion}

After introducing the connection between the thermodynamic
quantities corresponding to the black hole horizon and the
cosmological horizon, we give the effective thermodynamic quantities
of the RN-dS system, (\ref{eq2.4}), (\ref{eq4}) and (\ref{eq5}).
When describing the RN-dS system by the effective thermodynamic
quantities, it will exhibit a similar phase transition to Van der
Waals equation. In Sec.3 it shows that the position $x$ of the phase
transition point in RN-dS system is irrelevant to the electric
charge of the system. This indicates that for fixed charge when the
ratio of the black hole horizon and the cosmological horizon is
$x^c$, the second --order phase transition will occur. From Fig.1,
when the effective temperature $T_{eff} <T_{eff}^c $, the system
lies at a non-equilibrium state because of $\left( {\frac{\partial
P_{eff} }{\partial v}} \right)_{T_{eff} }
>0$ for some values of $v$. These states turn up at the small value of $v=r_c (1-x)$,
namely at the large value of $x>x^c$. This means that when the two
horizons are close to each other, the system is in non-equilibrium
state. Therefore the state in which the two horizons of RN-dS
approach does not exist. Only the states of RN-dS black holes with
$x<x^c$ can exist.

In Sec. 4 we analyzed the phase transition of RN-dS system. It shows
that at the critical point the specific heat at constant pressure,
the expansion coefficient $\beta $ and the compressibility $\kappa $
of the RN-dS system exist infinite peak, while the entropy and the
Gibbs potential G are continuous. Therefore for the phase transition
of the RN-dS system no latent heat and no specific volume changes
suddenly, it belongs to the second-order phase transition.

To understand black hole and cosmological singularities, or
distinguish all kinds of inflation models, or study the physics at
the Planck scale, specially to investigate the nature of the dark
energy which accounts for about $68.3\%$ of the substance of the
universe, a complete quantum theory of gravity is needed. Black
holes refer to gravity, quantum mechanics and thermodynamics, in
particular black holes in de Sitter space combine black holes with
cosmology. When considering the connection between the black hole
horizon with the cosmological horizon, it is possible to study the
non-equilibrium gravitational system, like RN-dS black hole. We are
looking forward to the research on the thermodynamic properties of
de Sitter space, such as phase transition and critical phenomena can
supply more information about quantum gravity and help to understand
the classic and quantum properties of de Sitter space.

\begin{acknowledgments}\vskip -4mm
This work is supported by NSFC under Grant
Nos.(11175109;11075098;11247261;11205097).
\end{acknowledgments}

\end{document}